\documentclass{article}

\usepackage{microtype}
\usepackage{graphicx}
\usepackage{subfigure}
\usepackage{booktabs} % for professional tables
\usepackage[subtle]{savetrees}
\usepackage{amssymb}

\usepackage{hyperref}

\usepackage{amsmath}
\usepackage{hyperref}

\usepackage[accepted]{icml2021}

\icmltitlerunning{Evaluating Fairness in Black-box Algorithmic Markets}

\begin{document}

\twocolumn[
\icmltitle{Evaluating Fairness in Black-box Algorithmic Markets: \\ A Case Study of Ride Sharing in Chicago}



\begin{icmlauthorlist}
\icmlauthor{Yuhan Liu}{pu}
\icmlauthor{Yuhan Zheng}{pu}
\icmlauthor{Siyuan Zhang}{pu}
\icmlauthor{Lydia T.~Liu}{pu}
\end{icmlauthorlist}

\icmlaffiliation{pu}{Department of Computer Science, Princeton University, Princeton, New Jersey, USA}

\icmlcorrespondingauthor{Yuhan Liu}{yuhanl@princeton.edu}
\icmlcorrespondingauthor{Lydia T.~Liu}{ltliu@princeton.edu}

\icmlkeywords{Algorithm auditing, gig economy, ML fairness}

\vskip 0.3in
]



\printAffiliationsAndNotice{}  % leave blank if no need to mention equal contribution

\begin{abstract}
This study examines fairness within the rideshare industry, focusing on both drivers' wages and riders' trip fares. Through quantitative analysis, we found that drivers' hourly wages are significantly influenced by factors such as race/ethnicity, health insurance status, tenure to the platform, and working hours. Despite platforms' policies not intentionally embedding biases, disparities persist based on these characteristics. For ride fares, we propose a method to audit the pricing policy of a proprietary algorithm by replicating it; we conduct a hypothesis test to determine if the predicted rideshare fare is greater than the taxi fare, taking into account the approximation error in the replicated model. Challenges in accessing data and transparency hinder our ability to isolate discrimination from other factors, underscoring the need for collaboration with rideshare platforms and drivers to enhance fairness in algorithmic wage determination and pricing.
\end{abstract}

\section{Introduction}
% The introduction should address the following:
% \begin{enumerate}
%     \item Problem statement: What problem does this paper solve? Or, what question does it answer?
%     \item Motivation: Why is this an important or challenging problem (question)? What is its societal significance?
%     \item Approach: How does this paper approach the problem? Introduce the data and methodologies.
%     \item Contributions/Novelty: What are the new contributions of this paper? Discuss conceptual contributions and technical contributions.
% \end{enumerate}

Rideshare platforms such as Uber and Lyft serve as algorithmic market intermediaries, matching drivers and riders \cite{chan2012ridesharing,agatz2010sustainable}. In addition to providing a matchmaking service, they determine the pricing for riders and the earnings for drivers for each trip, via a dynamic pricing strategy to manage the balance between rider demand and driver availability~\cite{shapiro2018between, pandit2019pricing, dong2021managing}. %Through this approach, prices are adjusted in real-time, increasing riders' fares and drivers' earnings through ``surge" pricing in areas with high demand at specific times~\cite{shapiro2018between, pandit2019pricing}. 
The algorithms driving this process optimize complex objectives, operating as black-box to both riders and drivers, which leads to large information asymmetry. Moreover, the algorithmic decisions regarding pricing and wages exacerbate the power imbalances~\cite{rosenblat2016algorithmic, yao2021together, jarrahi2019algorithmic, jarrahi2020platformic}. This power asymmetry undermines the emotional, physical, and financial well-being of both riders and drivers ~\cite{dubal2023algorithmic, woodcock2018gamification, watkins2023face}.

Given significant information and power imbalance, stakeholders of gig platforms have raised concerns regarding algorithm fairness~\cite{zhang2022algorithmic, hsieh2023designing}. While prior research has explored the individual drivers' perceptions and expectations of fairness through qualitative approaches, there remains a gap in quantitatively assessing these issues. This paper aims to fill that gap by analyzing the fairness of AI algorithms used in rideshare platforms, specifically focusing on their role in pricing and wage decisions. In this study, we employ publicly available data to measure the following: 1. Fairness in compensating drivers and 2. Fairness in charging riders.

We use the Chicago Transportation Network Provider Dataset, including the \href{https://data.cityofchicago.org/Transportation/Public-Passenger-Vehicle-Chauffeur-Survey-2021/uk68-3rjc/about_data}{Public Passenger Vehicle Chauffeur Survey}, \href{https://data.cityofchicago.org/Transportation/Transportation-Network-Providers-Trips-2023-/n26f-ihde/about_data}{Transportation Network Providers Trips}, and \href{https://data.cityofchicago.org/Transportation/Taxi-Trips-2013-2023-/wrvz-psew/about_data}{Taxi Trips} to measure the aforementioned areas. The Public Passenger Vehicle Chauffeur Survey Dataset is used to evaluate the compensation bias of drivers in terms of different aspects e.g. age, race, etc. The other two datasets are used to evaluate the fairness of the pricing algorithms of the rideshare platforms compared to the taxi companies in Chicago. 

The main contribution of this work is a quantitative, empirical investigation of algorithmic fairness in the two-sided market for ride-sharing, examining it from both the drivers' and the passengers' perspectives. We tackle a key challenge for performing a comparative audit of fares: given the AI algorithms are not open-sourced in most commercial settings, we attempt to replicate the pricing algorithm via machine learning in order to generate counterfactual fares, and account for the prediction error of the replicated algorithm in our hypothesis test. Finally, the study highlights the need for policy interventions to increase transparency in AI dynamic pricing algorithms used by rideshare platforms.

%\section{Related Work}
\paragraph{Related work}
%Algorithmic Decision Making of Wages and Prices in Gig Economy}
Recent research has revealed inequality in the gig economy across different genders, races, and socioeconomic statuses~\cite{hsieh2023designing}. Like traditional industries, the pay gap exists in crowdsourcing and freelancing platforms as prior work has shown~\cite{dubey2017analyzing, dunn2021gender, foong2021understanding}. In these platforms, compensation is often decided by service requesters~\cite{jarrahi2019algorithmic, jarrahi2020platformic}. However, in the rideshare industry, drivers' wages are determined by algorithms instead of humans, which raised people's attention to whether the pay gap still exists. Prior work by Cook and colleagues revealed a significant gender pay gap in Uber with the Chicago dataset~\cite{cook2021gender}. Our study aims to evaluate the fairness of drivers' wages by examining other demographic factors like age and gender, socioeconomic factors like insurance coverage, and working factors such as tenure to the platform and average working hours.

In addition to determining wages, algorithms also set the prices paid by riders. Unlike taxis, which use a transparent and fixed pricing strategy, ridesharing services have dynamic pricing influenced by trip distance, time, and demand-supply status~\cite{shapiro2018between, pandit2019pricing, chen2015peeking, hall2021pricing}. Prior works~\cite{ge2020racial, brown2018ridehail} have revealed discrimination in ridesharing, such as unequal wait times and cancellation rates across different demographic groups using simulated requests. Recent research also pointed out that price discrimination exists in different neighborhoods~\cite{pandey2021disparate}. According to recent reports, customers showed dissatisfaction with rideshare services and perceived them as overpriced ~\cite{nypost2023uberprices, missionlocal2021rideshare, vice2023farehike}. In this project, we aim to explore whether riders are fairly charged for the services they receive, addressing potential unfairness from the perspective of fair competition in the market.

%\paragraph{Algorithm Audit in Rideshare Platforms}
%Algorithm audit is a method of observing opaque algorithm output in order to draw conclusions about its inner workings and possible external impact~\cite{metaxa2021auditing}. 
% The opaqueness of the algorithm raised concerns regarding its fairness and societal impact, which drew attention from algorithm audit researchers~\cite{metaxa2021auditing, chen2015peeking}. In prior work, researchers used various methods for running audit studies, including examining different significance on existing datasets such as the Chicago datasets~\cite{cook2021gender, pandey2021disparate} and modeling algorithms by running regressions on simulated requests~\cite{chen2015peeking, ge2020racial, brown2018ridehail}. Our study is inspired by previous research, we ran Chi-square tests to examine the significant level of earnings differences between groups, and we attempted to replicate the pricing algorithm via machine learning to generate counterfactual fares.
\section{Data \& Method}
% The Data \& Methods section* should address the following:
% \begin{enumerate}
%     \item Describe the dataset that you will analyze, or collect. If you’re collecting the dataset, describe the method of data collection.
%     \item Why this dataset?
%     \item How will you use this dataset to answer the research question(s) that you have? 
% \end{enumerate}
\paragraph{Dataset}
We used the Public Passenger Vehicle Chauffeur Survey Dataset to evaluate fairness in compensating drivers. The dataset contains 7021 self-reported data points during 2021 in response to 73 questions. The survey asked about drivers' basic demographic information and also drivers' working experience, including their working schedule and average earnings. We filtered the dataset to exclude empty datapoints for each feature we were interested in. For most features, excluding education level and health insurance status, the filtered dataset has nearly 3000 non-empty self-reported data points. Both education level and health insurance status only have up to 1600 non-empty rows. 

For fairness in charging riders, we selected the Transportation Network Providers (TNP) Trips and Taxi Trips datasets. The TNP Trips monthly updates all the trips reported by rideshare companies, starting from Jan 1, 2023, to the current time, in Chicago. Each trip contains information such as pick-up/drop-off location, trip miles, trip length, trip fare breakdown, etc. We filtered out all carpool trips and all the trips after January 2024, resulting in \textbf{81,274,469} rows in the dataset. The Taxi Trips dataset contains data from 2013 to 2023 reported by taxi companies in Chicago. Each trip contains the same information as the TNP Trips Dataset. We filtered out all trips that happened before 2023 to make the time consistent, as well as trips reported in the invalid distance or incurring toll fees, resulting in \textbf{5,808,081} rows in the dataset.

%\subsection{Method}
\paragraph{Evaluating fairness in driver's compensation}
For fairness in drivers' compensation, we measure the bias in the following aspects regarding their self-reported earnings with statistical tests and visualization: 1. age of the driver, 2. race/ethnicity of the driver, 3. highest level of education of the driver, 4. health insurance status, 5. how long the driver has been driving for the platform, 6. on average how long the driver drives per day. 
% \vspace{-1.25em}
% \begin{itemize}
%     \item age of the driver
%     \vspace{-0.75em}
%     \item race or ethnicity of the driver
%     \vspace{-0.75em}
%     \item highest level of education of the driver
%     \vspace{-0.75em}
%     \item whether the driver holds health insurance
%     \vspace{-0.75em}
%     \item how long the driver has been driving for the platform
%     \vspace{-0.75em}
%     \item on average how long the driver drives per day
%     \vspace{-1.25em}
% \end{itemize}

We hypothesize that certain demographic factors, such as age, race, and level of education, may influence the compensation received by rideshare drivers on the platform. Age-related bias may be introduced to prefer older drivers for safety \cite{mccartt2009effects, guo2017effects}. Similarly, biases related to race may manifest through discriminatory user behavior, such as riders canceling trips or giving lower ratings based on the driver's race \cite{ge2020racial}. Regarding the level of education, discrimination may occur if the platform perceives drivers with higher education levels as being more competent or trustworthy, leading to preferential treatment and potentially higher earnings. Access to benefits such as health insurance could also contribute to disparities in earnings, as well as factors related to drivers' working conditions, such as the number of hours driven per day and part-time or full-time status. These hypotheses reflect concerns about potential biases within the rideshare platform, such as the possibility of favoring certain groups of drivers over others.

\paragraph{Evaluating fairness in customer's payment} %\label{sec:payment_fairness}
For fairness in charging riders, we make the assumption that the amount riders pay is the sum of the trip fare and additional charges (such as taxes, tolls, and service fees). Under this assumption, we evaluate only the pricing algorithm in rideshare platforms, assessing whether the trip fare charged is fair and reasonable compared to that of taxi companies. In other words, we do not consider the impact of additional charges like taxes, tolls, and service fees. We chose to compare with taxi companies because their pricing strategy is more transparent and fixed compared to that of rideshare platforms. By excluding the additional charges, we can also explore how dynamic or peak pricing impacts trip fares. Based on recent reports about the ``skyrocketed'' rideshare fares, we hypothesize that the rideshare platforms overcharged their riders. 

For fare charging, we use the distance of the trip (in miles), the time length of the trip (in seconds), the trip starting timestamp, and the pick-up/drop-off location as inputs, with the trip fare as the output, to replicate the algorithms used by the platforms to calculate fares. Building a replicated algorithm is necessary for this project because we do not (and no one does) have access to the algorithms and training data used by rideshare platforms. We then use the replicated algorithm to calculate how much each trip in the Taxi Trips dataset would cost if the trip were fulfilled on rideshare platforms. Subsequently, we conducted data analysis to interpret the results.

% \subsection{Limitations}
% There are several limitations of this study due to the limited information we have in the dataset. First, the data in the Public Passenger Vehicle Chauffeur Survey - 2021 Dataset is self-reported. Some information in the dataset is the survey respondents' estimation, which may be not accurate. Second, participation in the survey was optional, leading to possible sample bias, as rideshare drivers who chose not to participate might have done so for a variety of reasons, skewing the data representation. Furthermore, the replicated algorithms are based on pattern recognition of historical data and may not represent the ground truth. An assumption made is equating "other fees" with "tolls" due to insufficient details in the dataset. In reality, "other fees" could encompass a broader range of charges, which we have chosen to overlook. \\

% Despite these challenges, it's important to note that such limitations are not unique to this study and are often encountered in other rigorous research, Sampling bias, for instance, is a common issue in studies that rely on survey data. Given the privacy concerns, there was no accessible public dataset that offered a more accurate alternative to self-reported earnings for this research topic. Furthermore, the scenario in which "other fees" include costs other than tolls is expected to be infrequent, thereby minimizing its impact on our findings.

\section{Results}
\paragraph{Fairness in driver's compensation}
Our analysis found no significant difference in rideshare drivers' hourly wages based on age or education level. However, race/ethnicity, health insurance status, driver’s tenure, and weekly working hours significantly impacted wages. This conclusion is supported by Chi-square test results for each feature.
\vspace{-1em}
\begin{itemize}
    \item Race/ethnicity: Asian/Asian American drivers had the lowest median and mode hourly wage (\$12-13). The distribution of their wages was skewed towards lower categories, confirmed by a pairwise Chi-square test. Figure~\ref{fig:race} (displayed in Appendix~\ref{app:figures}) illustrates the distribution of wages in each reported race and ethnicity group.
    \vspace{-0.5em}
    \item Health Insurance Status: Drivers with Medicaid and Medicare had the lowest wages (median and mode: \$12-13 and $<$\$10, respectively). Their wage distribution was also skewed towards lower categories, with statistical significance confirmed by Chi-square tests (Figure \ref{fig:insurance} in Appendix~\ref{app:figures}).
    \vspace{-0.5em}
    \item Tenure: Drivers with less than 1 year or more than 10 years of tenure earned lower wages (mode: $<$\$10, median: \$12-13) as shown in Figure \ref{fig:tenure}, in Appendix~\ref{app:figures}. This finding contradicted our hypothesis that longer tenure correlates with higher wages as observed in other industries. The result of a pairwise Chi-square test demonstrated that the hourly wage distribution for drivers who stayed the shortest and the longest with the platform was statistically significant compared to drivers who lay in the middle.
    \vspace{-0.5em}
    \item Working Hours: Hourly wage increased as their weekly working hours increased (Figure \ref{fig:working_hours}, in Appendix~\ref{app:figures}). Our initial hypothesis was that when drivers drive longer, their earnings won’t increase due to gamification in the gig economy mentioned in previous literature~\cite{nagarajrao2024navigating, woodcock2018gamification}. However, our analysis shows proof against this hypothesis since there was a positive correlation between working hours and wage, and a pairwise Chi-square test showed statistical significance among different groups. 
     \vspace{-1em}
\end{itemize}

% \begin{figure}[t]
% %\vspace{-2em}
% \begin{center}
% \includegraphics[width=\linewidth]{figures/race.png}
% \end{center}
% \caption{
% Distribution of race groups in each hourly wage category. }
% \label{fig:race}
% \end{figure}

% \begin{figure}[t]
% \vspace{-0.5em}
% \begin{center}
% \includegraphics[width=\linewidth]{figures/insurance.png}
% \end{center}
% \caption{
% Distribution of insurance groups in each hourly wage category. }
% \label{fig:insurance}
% \end{figure}

% \begin{figure}[t]
%  %\vspace{-2em}
% \begin{center}
% \includegraphics[width=\linewidth]{figures/tenure.png}
% \end{center}
% \caption{
% Distribution of hourly wage category for each tenure group. }
% \label{fig:tenure}
% \end{figure}

% \begin{figure}[t]
% \vspace{-1em}
% \begin{center}
% \includegraphics[width=\linewidth]{figures/weekly_hour.png}
% \end{center}
% \caption{
% Distribution of hourly wage category for each working hour group. }
% \label{fig:working_hours}
% \end{figure}
Additionally, by examining the coefficients of the linear regression model that took all features above as inputs, we observed that race/ethnicity had the strongest effect on hourly wage. The effect magnitude of health insurance status, driver’s tenure and weekly working hours were similar and ranked just below the effect of race/ethnicity.

\begin{figure}[t]
% \vspace{-2em}
\begin{center}
\includegraphics[width=\linewidth]{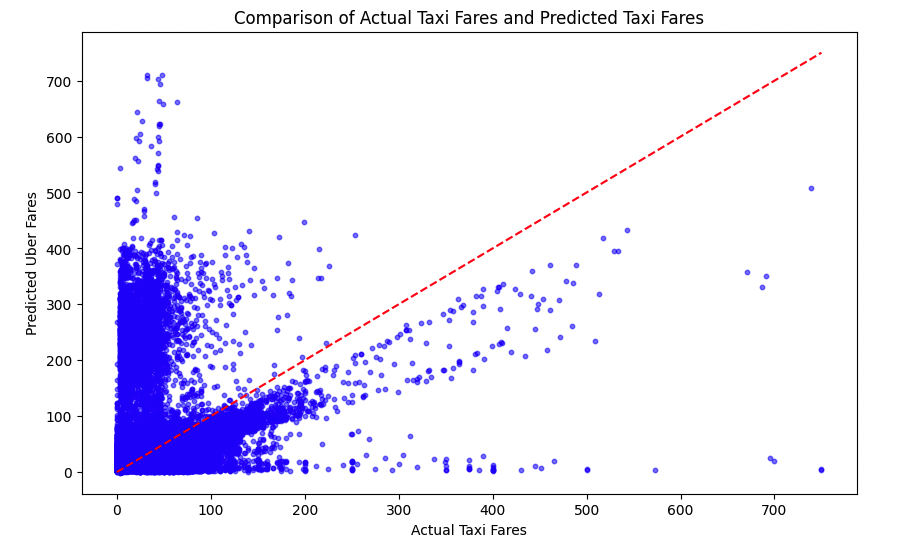}
\end{center}
\caption{
Comparison between the actual taxi fares and the predicted fares if trips were fulfilled by rideshare platforms. Each data point represents one trip. Points above the red dotted line indicate predicted fares are higher than actual fares, while points below indicate predicted fares are lower than actual fares.
}
\label{fig:actualvspredict}
\vspace{-1em}
\end{figure}

\paragraph{Fairness in customer's payment}
Consider the hypothesis that a customer is charged systematically differently for a ride taken by taxi vis-a-vis rideshare. A challenge to testing this hypothesis is that each ride is either taken via taxi or rideshare, but not both. Since each ride is different and may be more or less expensive due to its length and duration (even in the case of taxi rides), such differences must be taken into account. We take the following strategy:
For each ride $X$, we model \( f(X) \) as the rideshare fare and \( g(X) \) as the taxi fare. We train a model to predict the rideshare fare function, resulting in an approximation \( \hat{f}(X) \). Our best linear regression model had an RMSE of 5.35 and a $R^2$ of 0.70 when tested with the holdout test data from the TNP dataset. 
We take the taxi rides in our dataset \( \{X_1, X_2, \ldots, X_n\} \), making the assumption that it is a representative sample of rides, and compute the \emph{counterfactual} rideshare fares \( \hat{f}(X_i) \)'s (see Figure~\ref{fig:actualvspredict}). 
 An initial Mann–Whitney U test showed that there was a statistically significant difference between the distribution of predicted ride share trip fares and actual taxi fares (rejected the $H_{0}$ hypothesis with $p \text{ value}=0.000$). Further analysis revealed that 14.83\% of the 95\% confidence intervals for the counterfactual rideshare fares were below than actual taxi fares predicted, while 0.89\% were higher.

We test the hypothesis that the expected rideshare fare is greater than the taxi fare, with a one-tailed $t$-test. Specifically, the null hypothesis (\( H_0 \)) is that \( \mathbb{E}[f(X) - g(X)] \geq 0 \). %The alternative hypothesis (\( H_1 \)) is that \( \mathbb{E}[f(X) - g(X)] < 0 \), indicating that the expected rideshare fare is less than the taxi fare. 
 We calculate \( D_i = \hat{f}(X_i) - g(X_i) \), and its variance, taking into account the known variance of the error in \( \hat{f} \) in the calculation of the \emph{effective variance} of the t-statistic. Comparing the p-value to a significance level of 0.05, we failed to reject the null hypothesis that the difference between predicted rideshare fare and the actual fare is less than or equal to zero, suggesting insufficient evidence to conclude that rideshare platforms charged higher than taxi platforms. Our initial investigation shows that the differences in taxi and rideshare pricing policies, while significant, are nuanced; we may require more data than as well as further analysis to draw any stronger conclusions. Our approach of replicating an opaque algorithm from public data in order to perform hypothesis tests that are \emph{valid for the opaque algorithm} may be of independent interest.

% \paragraph{t-test}
\section{Discussion and Conclusion}
\paragraph{Disparities in drivers' compensation calls for policy intervention and more transparency}
We explore fairness issues regarding drivers' compensation from three perspectives: demographic features like age and race, other personal information like education level and health insurance, and work-related features like platform tenure and working time. Our results show that there is no significant evidence that age discrimination exists. %We attribute this to the difference between the traditional labor market and the gig economy. In gig platforms, the wage is determined by the algorithm, instead of human beings, so the source of unfairness is the training data~\cite{lee2023ai, birhane2024into}; in the process of building the wage determination algorithm, age is not taken into consideration. \footnote{Though we don't know exactly what features are used in algorithm training, we believe age, race, and ethnicity are not included as it is against \href{https://www.eeoc.gov/equal-paycompensation-discrimination}{Title VII}.}

However, our data analyses show a strong association between wage and race/ethnicity. Indirect human input like customer ratings can track racial bias and influence drivers' earnings, even if race/ethnicity is not directly used by pricing algorithms; more investigation is needed to understand if this constitutes wage discrimination. Previous work in traditional workplace
discrimination discussed the disadvantaged situation faced by African Americans and Hispanics~\cite{hirsh2010perceiving}. In the labor market with emerging technologies, the old problems are not eliminated but hidden in other forms. Given that there is no effective technical solution to mitigate hidden bias embedded in user input, negative impact could be mitigated through the platform setting up a compensation policy that balances bias in customer input and its influence on drivers' earnings to ensure the service quality while avoiding discrimination. 

Recent investigations into gig platforms have uncovered drivers' perceptions of unfairness, such as algorithms showing favoritism towards newer drivers~\cite{nagarajrao2024navigating} and potentially manipulating task allocations to create challenges in achieving certain Quests~\cite{nagarajrao2024navigating, krzywdzinski2021between}. Building upon previous work, we aligned our hypotheses with these findings. While our results indeed demonstrated a degree of algorithmic favoritism towards new drivers with some experience, we also observed a positive correlation between working time and wages. Interestingly, we found a misalignment between drivers' perceptions shown in previous literature and our quantitative findings, suggesting potential discrepancies in compensation fairness perceptions or limitations in our quantitative data's ability to capture crucial insights. Notably, our analysis relied on publicly available data, and further investigations are hindered by the lack of specific data. This underscores the critical need for enhanced algorithm transparency—ensuring not only comprehensibility for drivers from non-tech backgrounds but also facilitating data access for algorithm auditing researchers with technical expertise.

\paragraph{Transparency is needed to increase riders' perception of fairness towards dynamic pricing system}

The rideshare industry is not the first one, nor is it the only one that introduced a dynamic pricing mechanism. Other transportation industries like airplanes have been adopting similar pricing strategies for decades~\cite{proussaloglou1999choice}. However, while price differentiation exists across various transportation modes, the issue of price discrimination within rideshare platforms has garnered particular attention~\cite{pandey2021disparate}. Riders hold a biased impression that the rideshare service is overpriced~\cite{missionlocal2021rideshare, nypost2023uberprices, vice2023farehike}. Our findings indicate that when compared to taxi service — the most analogous transportation alternative to rideshare—around 15\% of rides result in lower expenditures for riders while the overcharging percentage is less than 1\%. We attribute the variance in perception between rideshare and the airline industry to two factors: on-demand nature and lack of transparency. 

In contrast to rideshare, the airline industry implements a comparable pricing model, adjusting fares based on supply, demand, and fuel costs. The primary distinction lies in transparency: flight ticket prices for various dates and times are readily available on airline websites, facilitating advance planning and coordination. Moreover, the abundance of data in the airline sector has spurred advancements in price prediction technologies~\cite{tziridis2017airfare, boruah2019bayesian}. The on-demand nature of rideshare services makes advance scheduling difficult and complicates price prediction. While AI algorithms are essential to these operations, their proprietary nature prevents full disclosure, leading to decreased rider trust. Our analysis highlights the importance of transparency in building both driver confidence and rider trust.

%Through quantitative analysis, we highlighted the difference between algorithmic management and traditional employment in terms of wages, pointing out the new fairness challenges in ``the future of work'' area. In addition, we also observed different findings which are in contrast to the previous literature. 

\paragraph{Conclusion and Future Work}
We investigated algorithmic fairness in the rideshare industry from the perspective of both drivers and riders and consequently raised concerns about discrimination in gig worker's labor market and transparency in these platforms' dynamic pricing systems, employing various statistical tests on public datasets. %With Chi-squared tests and statistics on wage for drivers with varying characteristics, 
We discovered that race/ethnicity, health insurance status, tenure, and working hours significantly impacted hourly wage. %While rideshare platforms rewarded people who worked for longer hours, they were potentially underpaying workers based on their race/ethnicity, health insurance status, and tenure. 
On the other hand, we revealed the difficulty of accessing fairness in customer's payment by replicating rideshare platform's pricing model with features that were traditionally used to calculate trip fares. %Our result showed that for 42\% of the trips tested, customers would be paying less than tradition taxi fare, while they would be paying more for the other 58\% of the trips. The cause of such price discrepancy could be solely the dynamic pricing mechanism or in combination with other hidden algorithmic biases. 
Given the limited amount of public datasets on worker's wages and limited studies on how drivers interact with rideshare platforms, we faced obstacles in isolating workforce discrimination from confounding factors. The lack of transparency in their pricing algorithm also posed challenges to justify variations in trip fares. %Future work may focus on working with rideshare platforms and the drivers' community to reduce the lack of data and transparency. With greater information access, we can gain a better understanding of fairness in algorithmically determined wages and pricing.

\nocite{langley00}

\bibliography{main}
\bibliographystyle{icml2021}

\newpage
\appendix
\onecolumn
\section{Figures}\label{app:figures}
\begin{figure}[hbt!]
%\vspace{-2em}
\begin{center}
\includegraphics[width=\linewidth]{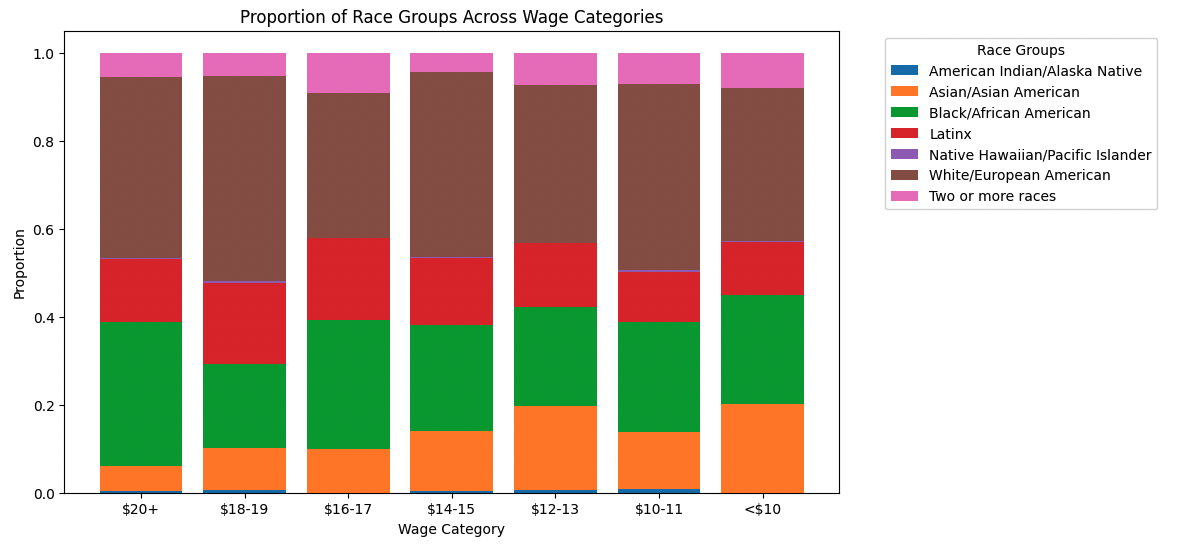}
\end{center}
\caption{
Distribution of race groups in each hourly wage category. }
\label{fig:race}
\end{figure}

\begin{figure}[hbt!]
\begin{center}
\includegraphics[width=\linewidth]{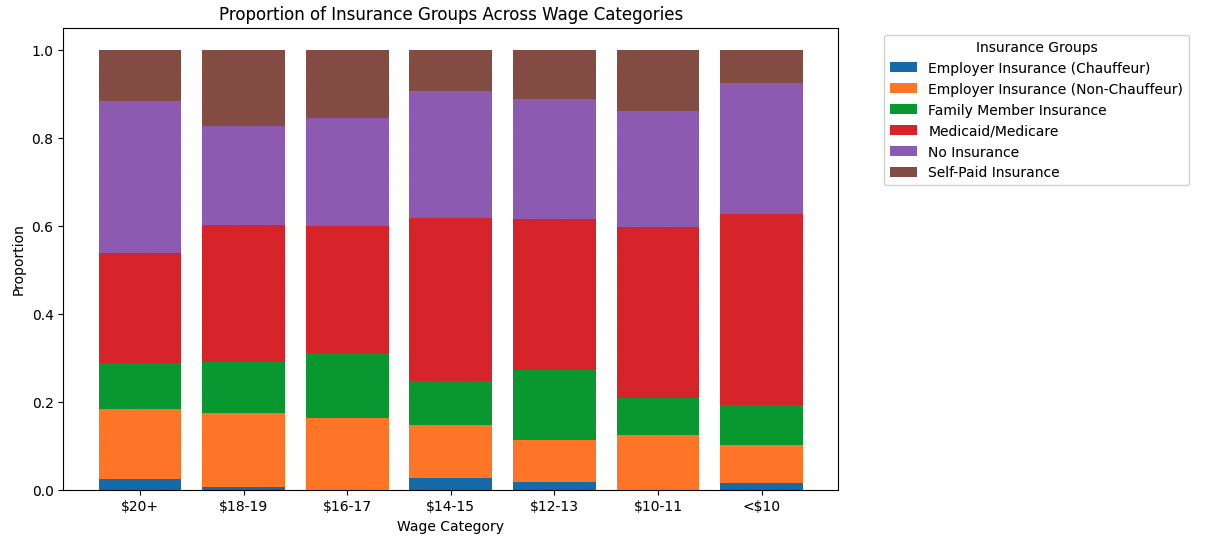}
\end{center}
\caption{
Distribution of insurance groups in each hourly wage category. }
\label{fig:insurance}
\end{figure}

\begin{figure}[hbt!]
 %\vspace{-2em}
\begin{center}
\includegraphics[width=0.7\linewidth]{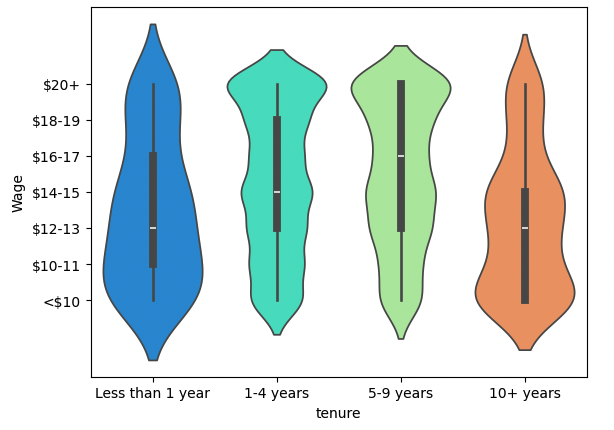}
\end{center}
\caption{
Distribution of hourly wage category for each tenure group. }
\label{fig:tenure}
\end{figure}

\begin{figure}[hbt!]
\vspace{-1em}
\begin{center}
\includegraphics[width=0.7\linewidth]{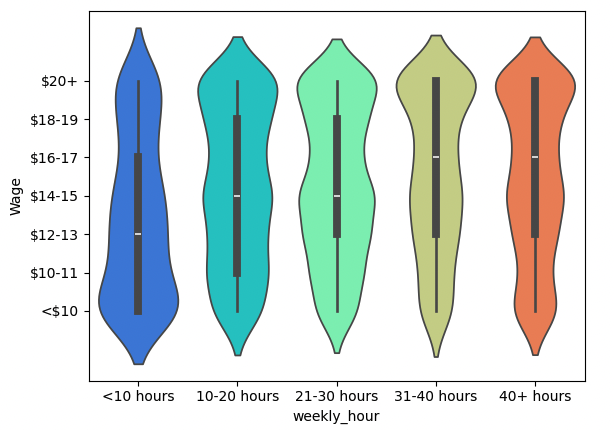}
\end{center}
\caption{
Distribution of hourly wage category for each working hour group. }
\label{fig:working_hours}
\end{figure}

\end{document}